\RequirePackage[l2tabu, orthodox]{nag}
\documentclass{article}

\usepackage[fleqn]{amsmath}
\usepackage[a4paper]{geometry}
\usepackage{graphicx}
\usepackage{microtype}
\usepackage{siunitx}
\usepackage{booktabs}
\usepackage[colorlinks=false, pdfborder={0 0 0}]{hyperref}
\usepackage{cleveref}

\usepackage{color}
\usepackage{cite}
\usepackage{parskip}
\usepackage{amssymb}
\usepackage{bm}
\usepackage[utf8]{inputenc}

\newcommand{\gdual}[1]{\overset{\:{}^{{}^{\boldsymbol{\sim}}}}{\smash[t]{#1}}} 
\newcommand{\dual}[1]{\overset{{}^{{}^{\boldsymbol{\sim}}}}{\smash[t]{#1}}} 
\newcommand{\gdualn}[1]{\overset{\:{}^{{}^{\boldsymbol{\neg}}}}{\smash[t]{#1}}} 
\newcommand{\dualn}[1]{\overset{{}^{{}^{\boldsymbol{\neg}}}}{\smash[t]{#1}}} 

\def\0{\mbox{\boldmath$\displaystyle\mathbb{O}$}}

\def\vp{\mbox{\boldmath$\displaystyle\boldsymbol{\varphi}$}}

\def\x{\mbox{\boldmath$\displaystyle\boldsymbol{x}$}}
\def\kb{\mbox{\boldmath$\displaystyle\boldsymbol{\kappa}$}}
\def\bz{\mbox{\boldmath$\displaystyle\boldsymbol{\zeta}$}}

\def\I{\openone}
\def\s{\mbox{\boldmath$\displaystyle\boldsymbol{\sigma}$}}
\def\p{\mbox{\boldmath$\displaystyle\boldsymbol{p}$}}
\def\e{\rm e}
\def\openone{\mathbb I}

\begin{document}
\date{\empty}

\noindent
{\textbf{\large Evading Weinberg's no-go theorem to construct mass dimension one fermions: Constructing darkness}}

\vspace{1pt}
\hspace{225pt}{\sc{Dharam Vir Ahluwalia}}\\

\noindent
{\textrm{Department of Physics}}\\
\textrm{Indian Institute of Technology Guwahati}\\
\textrm{Assam 981 039, India} \\

\noindent
{\textrm{Centre for Theoretical Physics}}\\
{\textrm{Jamia Millia Islamia,
\textrm{New Delhi 110025, India}}\\

\noindent
{\textrm{Theoretical Physics Division}}, 
{\textrm{Physical Research Laboratory}}\\
{\textrm{Ahmedabad 380 009, India}}\\

\noindent
\textrm{and}\\

\noindent
{\textrm{Centre for the Studies of the Glass Bead Game}}\\
\textrm{Chaugon, Bir, Himachal Pradesh 176 077, India}\\ 

\vspace{7pt}
\textcolor{red}{\hrule}

\vspace{51pt}
\begin{quote}
{\textbf{Abstract.}}  
Recent theoretical work reporting the construction of a new quantum field of spin one half fermions with mass dimension one requires that Weinberg's no go theorem must be evaded.
Here we show how this comes about. The essence of the argument is to first define a quantum field with due care being taken in fixing the locality phases attached to each of the expansion coefficients. The second ingredient is to systematically construct the 
dual of the expansion coefficients to define the adjoint of the field. The Feynman-Dyson propagator constructed from the vacuum expectation value of the field and its adjoint then yields the mass dimensionality of the field. For a quantum field constructed from a complete set of eigenspinors of the charge conjugation operator, with locality phases judiciously chosen, the Feynman-Dyson propagator determines the  mass dimension of the field to be one, rather than three halves. The Lorentz symmetry is preserved, locality anticommutators are satisfied, without violating fermionic statistics as needed for the spin one half field.
\end{quote}
\vspace{21pt}
Journal Reference: Europhys. Lett. 118 (2017) no.6, 60001\\
DOI: 10.1209/0295-5075/118/60001


\newpage

\noindent

Why do dark matter and dark energy~\cite{Clowe:2006eq,Bertone:2004pz,Copeland:2006wr,Peebles:2002gy,Ade:2013zuv,Komatsu:2010fb} have extremely limited interaction with the particles of the standard model of high energy physics? What type of quantum field theoretic formalism describes them~\cite{Padmanabhan:2002ji,Matos:1999et,Chung:1998bt,Guth:2014hsa,Detmold:2014qqa,Agarwal:2014oaa,Ahluwalia:2009rh,Basak:2012sn,daSilva:2014kfa}? The defining feature of the dark sector  is its `darkness' (that is, its limited or no interactions with the standard model field). Understanding this, or even constructing it theoretically, would dramatically accelerate our progress on the nature of physical reality and its building blocks. \textcolor{black}{In the $\Lambda$CDM cosmology,} there is much more, indeed more by a factor of roughly twenty  \textcolor{black}{(at the present epoch of cosmic evolution),} to existence then leptons, quarks, Higgs, and the standard model gauge bosons. 
 
 Despite undeniable 
success of the standard models of cosmology and high energy physics, it cannot be said with certainty if these problems suggest existence of new particles or if they hint at some incompleteness in the foundations of physics. If one takes the tentative view that dark fields
account for one or both of the dark sectors, then whatever these fields are they must be one representation or the other of the Lorentz algebra. At least locally, and at least in the low energy limit where Lorentz symmetries hold to a very high precision. Whether they may be broken in the dark sector, even in the present epoch, remains an open question~\cite{Ahluwalia:2010zn,Audren:2013dwa,Lee:2015tcc}. Here we shall confine  entirely to the realm of  unbroken Lorentz symmetries.

After the 1939 work of Wigner~\cite{Wigner:1939cj}, the first systematic effort to examine the particle content of the Lorentz algebra was undertaken by Weinberg. It spanned a series of papers published in the 1960s. The first of these, and most relevant for the present communication, was the 1964 paper~\cite{PhysRev.133.B1318}. It has now been expanded into its full textbook detail in the first few chapters of reference~\cite{Weinberg:1995mt}. Among other things these chapters construct Dirac's quantum field from first principles of relativity and quantum mechanics. This is done without invoking Dirac equation and the work results in a no go theorem on the impossibility of constructing another spin one half quantum field without violating Lorentz symmetries, and locality. Here we briefly review the relevant aspects of this no go theorem and show that it can be evaded to  construct fundamentally new type of quantum fields with the potential to shed light on the darkness of the dark sector.

The kinematic structure associated with the spin one half fermions of the standard model of high energy physics is contained in $\Psi(x)$, the Dirac quantum field. Assuming a Lorentz invariant vacuum and considering the space-time transformation properties of single particle states, exploiting cluster decomposition principle and invariance of the S-matrix (along with covariance under parity), Weinberg arrives at
\begin{equation}
\Psi(x) =  \int \frac{d^3\p}{(2\pi)^3} \sqrt{\frac{m}{E(\p)}} \sum_{\sigma=\pm{1/2}}\left[ u(\p,\sigma) e^{- i p^\mu x_\mu} a(\p,\sigma) + v(\p,\sigma) e^{ i p^\mu x_\mu} b^\dagger(\p,\sigma) \right] \label{eq:Psi}
\end{equation} 
from which, without reference to any wave equation or a Lagrangian density, the Feynman-Dyson propagator follows\footnote{We use the space-time metric with diagonal $\{1,-1,-1,-1\}$ and all symbols have their usual meaning. For instance, in (\ref{eq:FD-prop-Dirac}) $\mathfrak{T}$ represents time ordering operator in its usual sense.}
 \begin{equation}
S^{\mathrm{Dirac}}_{\mathrm{FD}}(x^\prime- x)  =  \left\langle\hspace{4pt}\left\vert \mathfrak{T} \left( \Psi(x^\prime) \overline{\Psi}(x)\right)\right\vert\hspace{4pt}\right\rangle
=  i    \int\frac{\text{d}^4 p}{(2 \pi)^4} 
{\e}^{- i p^\mu \left(x^{\prime}_\mu - x_\mu\right)} \left[  \frac{ \gamma_\mu p^\mu + m\I}{p_\mu p^\mu - m^2 + i \epsilon} \right] \label{eq:FD-prop-Dirac}
\end{equation}
where $\overline{\Psi}(x) \stackrel{\mathrm{def}}{=} \Psi^\dagger(x)\gamma^0$.
The dual of the expansion coefficients $u(\p,\sigma)$ and  $v(\p,\sigma)$ is
given by $\overline{u}(\p,\sigma) = u^\dagger(\p,\sigma) \gamma^0$, and  
$\overline{v}(\p,\sigma) = v^\dagger(\p,\sigma) \gamma^0$. The creation and the annihilation operators satisfy the fermionic statistics. In the Weinberg formalism, the explicit expressions for the expansion coefficient at rest, $\p =\bm{0}$, are derived. These are not taken as `solutions at rest' of any wave equation. After a discussion spreading over some two hundred pages Weinberg find these to be 
\begin{subequations}
\begin{equation}
u(\bm{0},1/2) = \frac{1}{\sqrt{2}}\left[\begin{array}{c} 1 \\ 0 \\ 1 \\ 0
\end{array}\right],\quad u(\bm{0},-1/2) = \frac{1}{\sqrt{2}}\left[\begin{array}{c} 0 \\ 1 \\ 0 \\ 1
\end{array}\right]\label{eq:u0}
\end{equation}
and
\begin{equation}
v(\bm{0},1/2) = \frac{1}{\sqrt{2}}\left[\begin{array}{c} 0 \\ 1 \\ 0 \\ -1
\end{array}\right],\quad v(\bm{0},-1/2) = - \frac{1}{\sqrt{2}}\left[\begin{array}{c} 1 \\ 0 \\ -1 \\ 0
\end{array}\right] .\label{eq:v0}
\end{equation} 
\end{subequations}
The boost operator for the expansion coefficients, again without a reference to a wave equation or a Lagrangian density, follows from the left- and right- Weyl representations of the Lorentz symmetries and reads
\begin{equation}
\sqrt{\frac{E+m}{2 m}}\left(\begin{array}{cc}
 \I + \frac{\s\cdot\p}{E+m} & \0 \\
\0 & \I - \frac{\s\cdot\p}{E+m}
\end{array}\right)
\end{equation}
It is simply an exponentiation, $\exp(i \kb\cdot\vp)$, of the boost generator found in equation~(\ref{eq:pi}) below ($\vp$~is the boost parameter).
Its action on the expansion coefficients at rest, (\ref{eq:u0}) and (\ref{eq:v0}),  yields $u(\p,\pm1/2)$ and $v(\p,\pm1/2)$ that appear in (\ref{eq:Psi}). The momentum space Dirac operator, $\gamma_\mu p^\mu \pm m \I$, is first seen in the spin sums, $\sum_\sigma  u(\p,\sigma)\overline{u}(\p,\sigma)$ and $\sum_\sigma  v(\p,\sigma)\overline{v}(\p,\sigma)$ that arise in evaluating the Feynman-Dyson propagator, (\ref{eq:FD-prop-Dirac}). It is thus that for $\Psi(x)$, Dirac's $(i\gamma_\mu\partial^\mu - m\I)$ enters the Lagrangian density and not the Klein Gordon operator, resulting in the mass dimension of $\Psi(x)$ to be $3/2$ and not $1$. The Dirac quantum field yields $S_{\mathrm{FD}}(x^\prime- x) = \left\langle\hspace{4pt}\left\vert \mathfrak{T} \left( \Psi(x^\prime) \overline{\Psi}(x)\right)\right\vert\hspace{4pt}\right\rangle $ as the Dirac propagator, and not the Klein Gordom propagator, even though $\Psi(x)$ is annhilated by the Dirac as well as the Klein Gordon operator.

First, it is to be noted that the correct pairing of the creation and annihilation operators  in~(\ref{eq:Psi}) requires that expansion coefficients be identified as the boosted form of (\ref{eq:u0}) and  (\ref{eq:v0}).
 In addition, the phase associated with each of the rest spinors is not arbitrary, to be simply set to unity, but these `locality' phases must be derived. Weinberg formalism yields 
both of these results without ambiguity. 
 It is also to be noted that references~\cite{Ryder:1985wq,Hladik:1999tt} contain an important error in the derivation of Dirac equation~\cite{Ahluwalia:1998dv,Gaioli:1998ra}. These errors, in effect, project out antiparticles. But then through two compensating mistakes 
 the authors of the mentioned works recover the correct equation in configuration space:
\textcolor{black}{
 \begin{quote}
 In the rest frame the right-transforming and left-transforming components of the Dirac spinors
 carry relative phases of $+1$ and $-1$ respectively for the particle and antiparticle spinors -- this is evident from (\ref{eq:u0}) and (\ref{eq:v0}). When derivations of Dirac equation given in~\cite{Ryder:1985wq,Hladik:1999tt} assume $\phi_R(\textbf{0})=\phi_L(\textbf{0})$ then the neglect of the  $\phi_R(\textbf{0})=- \phi_L(\textbf{0})$ results in an error and only yields mommentum-space Dirac equation for the particle spinors and not the antiparticle spinors. The naive replacement $p_\mu \to i\partial_\mu$ works for the particle spinors when acting on the configuration space representation through the following detail $i \partial_\mu u(\textbf{x}) = i (-i p_\mu) u(\textbf{x})$, but fails if one neglects the momentum-space Dirac equation resulting from $\phi_R(\textbf{0})=- \phi_L(\textbf{0})$ for which the mass terms differs by a sign (that is, instead of getting $(\gamma^\mu p_\mu -m) u(\textbf{p}) = 0$ one then obtains $(\gamma^\mu p_\mu  + m) v(\textbf{p}) = 0$). The result $i \partial_\mu v(\textbf{x}) = i (i p_\mu) v(\textbf{x})$, then yields $(i\gamma^\mu \partial_\mu - m) \psi(\x) = 0$. The error of two signs happens in assuming the same sign of the mass term for the $u(\textbf{p})$ and  $v(\textbf{p})$ spinors, and it is compensated by neglecting the sign differences in the action of $i \partial_\mu$ on these spinors (in the configuration space).
 \end{quote}
}

We have gone into this digression only to emphasize the elegance, care, and power of Weinberg's construction of quantum fields and how easily important errors have propagated in literature on the description of spin one half fermionic field. Another error that has propagated a large fraction of quantum field theory texts about the Dirac field is in the pairing of the expansion coefficients with the annihilation and creation operators. In its simplest version, apart from the locality phases, the error constitutes in the misidentification of the expansion coefficients $v(\bm{0},\pm 1/2)$ as $v(\bm{0},\mp  1/2)$ -- see, for example, Ryder's and Folland's monographs on quantum field 
theory~\cite{Ryder:1985wq,Folland:2006gb}.

Second, we raise the possibility that one can indeed evade the Weinberg no go theorem on the uniqueness of Dirac's, modulo the 1937 observation of Majorana~\cite{Majorana:1937vz}, field. The new observation is the following: 
\begin{quote}
The dual of the expansion coefficients defined immediately after (\ref{eq:FD-prop-Dirac}) is not unique. In fact, if the expansion coefficients are taken as the eigenspinors of the charge conjugation operator their norm under the Dirac dual identically vanishes. In reference~\cite[Appendix P]{Aitchison:2004cs} a lack of appreciation of this fact leads to abandoning an attempt to construct a Lagrangian density for c-number Majorana spinors. 
\end{quote}
And since the Feynman-Dyson propagator crucially depends on the expansion coefficients and their dual a crevice opens that, in principle, can evade the Weinberg no go theorem with important physical implications. To explore the crevice we consider, as an example, a quantum field with its expansion coefficients taken as self and antiself conjugate eigenspinors of the charge conjugation operator~\cite{Ahluwalia:2016rwl}
	\begin{equation}
	\mathfrak{f}(x) \stackrel{\mathrm{def}}{=} \int \frac{\text{d}^3p}{(2\pi)^3}  \frac{1}{\sqrt{2 m E(\p)}} \sum_{\alpha=\pm} \left[ \lambda^S(\p,\alpha)  e^{- i p^\mu x_\mu} c(\p,\alpha)
	+\, \lambda^A(\p,\alpha)  e^{ i p^\mu x_\mu} d^\dagger(\p,\alpha)\right]
	\end{equation}
The pairing of the expansion coefficients with the annihilation and creation operators (which are fermionic), and  locality phases [the counterpart of phases arrived at in
(\ref{eq:u0}) and (\ref{eq:v0})], is taken as in~\cite{Ahluwalia:2016rwl}. As noted above, it is readily verified that under the Dirac dual the norm of both the $\lambda^S(\p,\alpha)$ and $\lambda^A(\p,\alpha)$ identically vanishes; that is:   
	\begin{equation}
	\overline{\lambda}^S(\p,\alpha) \lambda^S(\p,\alpha) = 0 = \overline{\lambda}^A(\p,	\alpha) \lambda^A(\p,\alpha)
	\end{equation}
In the sense made precise below, we construct the duals of $\lambda^S(\p,\alpha)$ and $\lambda^A(\p,\alpha)$ by demanding not only the Lorentz invariance of the bilinear invariants but also the Lorentz invariance of the spin sums. 
We proceed in two steps.
To implement the first of the two constraint we introduce an `intermediate' dual
\begin{equation}
\widetilde{\lambda}(\p,\alpha) \stackrel{\mathrm{def}}{=} \big[\Xi(\p) \lambda(\p,\alpha)\big]^\dagger \eta \label{eq:dual}
\end{equation}
with~\cite{Ahluwalia:2016rwl,Speranca:2013hqa}
\begin{align}
		\Xi(\p)   \stackrel{\rm def}{=}  \frac{1}{2 m} \sum_{\alpha}	
			  \Big[\lambda^S(\p,\alpha)\overline\lambda^S(\p,\alpha) 	 
			  - \lambda^A(\p,\alpha)\overline\lambda^A(\p,\alpha) 	
			  \Big].		\label{eq:Xi}
	\end{align}
The sole task of the $\Xi(\p)$ is to implement the mapping  $\lambda^{S,A}(\p,\pm) \leftrightarrow  \lambda^{S,A}(\p,\mp)$ up to a constant phase factor; whose value is encoded in $\Xi(\p)$. The constraint of invariance under the boosts requires $\eta$ to anti-commute with the generator of the boosts in the $(1/2,0)\oplus(0,1/2)$ representation space, while counterpart of this constraint under rotations requires $\eta$ to commute with the 
generator of rotations
\begin{equation}
\left\{\eta,\kb_i\right\} = 0,\quad\left[\eta,\bz_i\right] = 0,\qquad i=x,y,z \label{eta-zimpok9}
\end{equation}
where the boost and rotation generators are given by
\begin{equation}
				\kb  = \left[
				\begin{array}{cc}
				- i \s/2 & \0\\
					\0 & + i \s/2
					\end{array}
					\right], \quad
 \bz = \left[
\begin{array}{cc}
 \s/2 & \0\\
\0 &  \s/2
\end{array}
\right].  \label{eq:pi}
\end{equation}
It happens because $\kb^\dagger = -\kb$ and $\bz^\dagger = \bz$.
Requiring the norm $\widetilde{\lambda}(\p,\alpha) {\lambda}(\p,\alpha^\prime)$ to be real, the constraints (\ref{eta-zimpok9}) leads $\eta$ to have the form
\begin{equation}
\eta = \left(\begin{array}{cccc}
0 & 0 & a & 0\\
0 & 0 & 0 & a \\
b & 0 & 0 & 0\\
0 & b & 0 &0
\end{array}\right),\qquad a,b \in \Re .\label{eta-ab}
\end{equation}
The invariance of the norm under parity forces $a$ and $b$ to be equal~\cite{Ahluwalia:2016rwl}. This reduces $\eta$ to 
$
\eta = a \gamma^0.
$ 
The $a$ is now simply a normalization factor and we set it to unity to yield $\eta = \gamma^0$. If the same exercise is carried out with the Dirac spinors then the Dirac dual is seen to result from $\Xi(\p) = \I$.
The definition (\ref{eq:dual}), and expression for $\Xi(\p)$ in (\ref{eq:Xi}), 
yields the following orthonormality relations
\begin{subequations}
\begin{align}
& \widetilde\lambda^S(p^\mu,\alpha) \lambda^S(p^\mu,{\alpha^\prime})
 = 2 m \delta_{\alpha\alpha^\prime},\quad  
   \widetilde\lambda^A(p^\mu,\alpha) \lambda^A(p^\mu,{\alpha^\prime})
 = - 2 m \delta_{\alpha\alpha^\prime}  \label{eq:onas}\\
 & \widetilde\lambda^S(p^\mu,\alpha) \lambda^A(p^\mu,{\alpha^\prime})
 = 0 = \widetilde\lambda^A(p^\mu,\alpha) \lambda^S(p^\mu,{\alpha^\prime})
\label{eq:onsaas}
\end{align}
\end{subequations}
These are manifestly Lorentz invariant. However, the spins sums
$\sum_\alpha \lambda^S(\p,\alpha) \widetilde{\lambda}^S(\p,\alpha)$ and
$\sum_\alpha \lambda^A(\p,\alpha) \widetilde{\lambda}^A(\p,\alpha)$ are found to be covariant only under a subgroup of Lorentz~\cite{Ahluwalia:2010zn}. This circumstance asks us to look for a freedom in the definition of the introduced dual so that the orthonormality relations preserve their Lorentz invariance but which makes the spin sums also Lorentz invariant.

To explore this possibility we consider the following modification to the definition of the introduced dual
\begin{equation}
\widetilde{\lambda}^S(\p,\alpha) \to \gdualn{\lambda}^S(\p,\alpha) \stackrel{\mathrm{def}}{=} \dual{\lambda}^S(\p,\alpha) \mathcal{A},\quad
\widetilde{\lambda}^A(\p,\alpha) \to \gdualn{\lambda}^A(\p,\alpha) \stackrel{\mathrm{def}}{=} \dual{\lambda}^A(\p,\alpha) \mathcal{B} 
\end{equation}
with $\mathcal{A}$ and $\mathcal{B}$ constrained to have the 
following non-trivial properties: the $\lambda^S(\p,\alpha)$ must be  eigenspinors of $\mathcal{A}$ with eigenvalue unity,  and similarly $\lambda^A(\p,\alpha)$ must be  eigenspinors of $\mathcal{B}$ with eigenvalue unity
\begin{subequations}
\begin{equation}
\mathcal{A} \lambda^S(\p,\alpha) = \lambda^S(\p,\alpha),\quad
\mathcal{B} \lambda^A(\p,\alpha) = \lambda^A(\p,\alpha).\label{eq:4jan-a}
\end{equation}
In addition, $\mathcal{A}$ and $\mathcal{B}$ must be such that
\begin{equation}
\gdual{\lambda}^S(\p,\alpha)\mathcal{A} \lambda^A(\p,\alpha^\prime)=0,\quad
\gdual{\lambda}^A(\p,\alpha)\mathcal{B} \lambda^S(\p,\alpha^\prime)= 0.
\label{eq:4jan-b}
\end{equation}
\end{subequations}
 If $\mathcal{A}$ and $\mathcal{B}$ exist satisfying the just stated constraints,
the new dual would leave the orthonormality relations (\ref{eq:onas})-(\ref{eq:onsaas}) unaltered in form
\begin{subequations}
\begin{align}
& \gdualn\lambda^S_\alpha(p^\mu) \lambda^S_{\alpha^\prime}(p^\mu)
 = 2 m \delta_{\alpha\alpha^\prime},\quad
  \gdualn\lambda^A_\alpha(p^\mu) \lambda^A_{\alpha^\prime}(p^\mu)
 = - 2 m \delta_{\alpha\alpha^\prime} \label{eq:zimpokJ9bn} \\
 &  \gdualn\lambda^S_\alpha(p^\mu) \lambda^A_{\alpha^\prime}(p^\mu) = 0 =
 \gdualn\lambda^A_\alpha(p^\mu) \lambda^S_{\alpha^\prime}(p^\mu)\label{eq:zimpokJ9cn}
\end{align}
\end{subequations}
but the same very  re-definition would alter the spin sums. The task then would be to see if this can be accomplished in such a way that the spins sums become Lorentz invariant. It turns out that the exercise can be done only if one multiplies the Lorentz-violating piece in the spin sums by a parameter $\tau\in\Re$ and then taking the $\tau\to 1$ limit~\cite{Ahluwalia:2016rwl}. This exercise reveals that no solution exists at $\tau=1$, but a solution does exist in the infinitesimally close neighborhood of  $\tau=1$. We take this as a physically acceptable solution, and find that the redefinition of the dual does yield a Lorentz invariant set of spin sums 
	\begin{align}
	\sum_\alpha \lambda^S(\p,\alpha) \gdualn{\lambda}^S(\p,\alpha) = 2 m
	 \I ,\quad
	\sum_\alpha \lambda^A(\p,\alpha) \gdualn{\lambda}^A(\p,\alpha) = - 2 m. \I
	\label{eq:spinsums-SA}
	\end{align}
We thus introduce the adjoint of $\mathfrak{f}(x)$
	\begin{equation}
	\dualn{\mathfrak{f}}(x) \stackrel{\mathrm{def}}{=} \int \frac{\text{d}^3p}{(2\pi)^3}  	\frac{1}{ 	\sqrt{2 m E(\p)}} \sum_\alpha \left[ \gdualn\lambda^S(\p,\alpha)  
	e^{ i p^\mu x_\mu} c(\p,\alpha)
	+\, \gdualn\lambda^A(\p,\alpha)  e^{- i p^\mu x_\mu} d^\dagger(\p,\alpha)\right]
	\end{equation}
and, using  (\ref{eq:spinsums-SA}), evaluate the Feynman-Dyson propagator
\begin{align}
S_{\text{FD}}(x^\prime- x)  &=  \left\langle\hspace{4pt}\left\vert \mathfrak{T} \left( \mathfrak{f}(x^\prime) \gdualn{f}(x)\right)\right\vert\hspace{4pt}\right\rangle\\
&=     \int\frac{\text{d}^4 p}{(2 \pi)^4} 
{\e}^{- i p^\mu \left(x^{\prime}_\mu - x_\mu\right)} \left[  \frac{ \openone_4}{p_\mu p^\mu - m^2 + i \epsilon} \right] \label{eq:FD-prop-b}
\end{align}
\textcolor{black}{It satisfies:
$\left(\partial_{\mu^\prime}   \partial^{\mu^\prime}  + m \I_4    \right) S_{\text{FD}}(x^\prime- x)  =
-\delta^4(x^\prime-x)$.}

This is a remarkable result: It gives us a spin one half fermionic field with mass dimension one. The Lagrangian density now readily follows, and from that one can calculate the canonically conjugate momentum and  ascertain  that the new field satisfies the locality anticommutators. The mass dimensionality of the new field has a mismatch with the standard model fermions and thus it cannot enter the standard model doublets. In the process the new field, because of its mass dimensionality, becomes a first-principle candidate for self-interacting dark matter. The self interaction arises from the dimension four quartic self interaction 
$g \big[\gdualn{\mathfrak{f}}(x) \mathfrak{f}(x)\big]^2$ with $g$ a dimensionless coupling constant. The new dark particles associated with $\mathfrak{f}(x)$ may  be detected through a dimension four coupling of the new field with the Higgs doublet, $g^\prime  
\gdualn{\mathfrak{f}}(x) \mathfrak{f}(x) \phi^\dagger(x)\phi(x) $; where $g^\prime$ is another dimensionless coupling.

It is also worth noting that the expansion coefficients $\lambda^{S,A}(\p,\alpha)$ of the new field do not satisfy Dirac equation, but only Klein Gordon equation~\cite{Ahluwalia:2016rwl}. The Weinberg no go theorem is evaded, and with that opens up a fertile possibility to construct other fields relevant for the dark sector and to provide us new insights into the standard model. One such possibility arises when one considers the $(1/2,1/2)$ representation space. Again there emerges a freedom in the construction of the dual. This time it appears as a phase factor multiplying the usual space-time metric, with significant implications for understanding the deeper origin of Higgs-like particles. The issue of the $\tau$ deformation is specific to 
$\mathfrak{f}(x)$ and is not generic. 

By evading the Weinberg no go theorem, and by understanding the underlying general structure of duals and adjoints, we find a whole new range of possibilities to extend the standard model of high energy physics. Many of these new constructs shall carry a natural darkness, and would be endowed with new properties with respect to the parity, charge conjugation, and time reversal. Mass dimension one fermions are only a first concrete example of this new programme. 

\textcolor{black}{
As is evident from the vast literature on the subject various classes of spinors can be defined. Some of these, for instance, depend on the bilinear covariants and the relevant duals, while others depart from the restriction on zero curvature of the spacetime manifold, while still other depend on the 
behaviour of the spinors under the symmetries of charge conjugation and parity
(see, for example, \cite{Cavalcanti:2014wia,Fabbri:2016laz,Wigner:1962ep}). Whether all of these can serve as expansion coefficients to construct local quantum fields remains open.
}


\providecommand{\href}[2]{#2}\begingroup\raggedright\endgroup

\end{document}